\documentstyle[preprint,aps,epsf]{revtex}
\begin{document}
\draft
\title{Nuclear-polarization effect to the hyperfine structure in heavy
multicharged ions}

\author{A.V.~Nefiodov $^{\rm a,b}$, G.~Plunien $^{\rm a}$, and G.~Soff 
$^{\rm a}$} 
\address{
$^{\rm a}$ Institut f\"ur Theoretische Physik, Technische  Universit\"at
Dresden, Mommsenstra{\ss}e 13, D-01062  Dresden, Germany  \\ 
$^{\rm b}$ Petersburg Nuclear Physics Institute, 188300 Gatchina,
St.~Petersburg, Russia}

\date{Received \today}
\maketitle

\widetext
\begin{abstract}
We have investigated the correction to the hyperfine structure of heavy
multicharged ions, which is connected with the nuclear-polarization effect
caused by the unpaired bound electron. Numerical calculations are performed
for hydrogenlike ions taking into account the dominant collective nuclear
excitations. The correction defines the ultimate limit of precision in
accurate theoretical predictions of the hyperfine-structure splittings.
\end{abstract}
\pacs{PACS numbers: 12.20.Ds, 31.30.Gs, 32.10.Fn, 32.10.Dk}

\narrowtext
During the last decade a significant progress has been achieved in
investigations of the hypefine structure (hfs) in heavy multicharged ions. To
date, accurate measurements of hfs splittings were performed for a number of
elements by using optical spectroscopy at the experimental storage ring at GSI
in Darmstadt and at the Super-EBIT at the Lawrence Livermore National
Laboratory \cite{Kl94,JRCL96,JRCL98,PB98,PSe98,PB01}. One of the  major
purposes of these investigations consists in testing nontrivial effects of
bound-state QED in intense nuclear fields with an accuracy on the level of a
few percent. Experiments on the hfs splitting allow one also to probe internal
nuclear structure, in particular, the  magnetic moment distribution within the
nucleus (the so-called Bohr-Weisskopf effect). Though a rough estimate of the
latter effect can be obtained in the framework of the single-nucleon model,
its accurate calculation with the use of a microscopic nuclear theory still
has to be performed. 

Two attempts to calculate the extended magnetization distribution taking into
account nuclear many-body corrections should be mentioned here. The first
approach is based on the dynamic correlation model \cite{Tom88}, which allows
to include the excited-core configurations in the nuclear wave function.
Numerical calculations performed in the framework of this model provide 
accurate values for the nuclear magnetic moments. However, no satisfactory 
agreement with experimental results for the hfs splittings has been achieved 
\cite{Tom02}. The second approach, which also takes into account the nuclear
core polarization by the unpaired nucleon, is based on the theory of finite
Fermi systems \cite{SDm02}. In this case, the correction due to the nuclear
magnetization distribution calculated for hydrogenlike bismuth turned out to
be too small compared with the corresponding value deduced from the experiment.

At present it appears to be unlikely to obtain an accurate theoretical value
for the Bohr-Weisskopf correction. However, one can eliminate it to a large
extend in combined measurements of the hfs splittings in H- and Li-like ions
\cite{Shab98}. For the ground state of ${}^{209}$Bi${}^{80+}$, two independent
calculations have lead to similar predictions, $797.1(2)$ meV \cite{Sh00} and
$797.15(13)$ meV \cite{SSc01}, respectively. While both theoretical results
agree with the experimental value of $820(26)$ meV \cite{PB98}, further
efforts to measure the predicted splitting with higher precision have not been
successfull. 

In the present paper we evaluate a correction, which has not been
previously considered in calculations of the hyperfine structure. The
correction is connected with the nuclear-polarization effect caused by the
unpaired bound electron. While the corresponding contribution is relatively
small in the case of hydrogenlike ions, it turns out to be non-negligible for
accurate theoretical predictions of the hfs splittings in Li-like heavy ions.
Since the uncertainty of our calculation is comparable with the magnitude of
the nuclear-polarization effect itself, the latter cannot be completely
eliminated by extracting the Bohr-Weisskopf contribution in accordance with 
Shabaev's idea \cite{Shab98}. The effect under consideration sets a natural
limit up to which one can test bound-state QED, even if a specific difference
of the hfs splittings of H- and Li-like ions is introduced \cite{ShAN01}.

In the following, we shall consider a hydrogenlike ion with a nonzero-spin
nucleus, so that the total angular momentum of an atom $\bbox{F}$ is defined
by the coupling of the nuclear spin $\bbox{I}$ with the total angular momentum
of an electron $\bbox{j}$. The hfs energy shift due to the magnetic-dipole
interaction is given by
\begin{equation}
\Delta E_{\mu}(F)=\frac{1}{2}\left[ F(F+1) -I(I+1) -j(j+1)\right]
A_{n\kappa} . 
\label{eq1}                                        
\end{equation} 
Here $A_{n\kappa}$ is the hfs constant, which depends on the electron state
characterized by the standard set of quantum numbers. For electron states
with $j=1/2$, the hfs splitting between the levels with $F=I+1/2$ and $F=I-1/2$
is just $\Delta E_{n\kappa} = (I+1/2)A_{n\kappa}$. Since the nuclear size is
rather small with respect to the radius of the electron orbit, the
hyperfine structure can be fairly understood in the framework of the
external-field approximation. This allows to treat the magnetic field of a
nucleus as a perturbing potential in calculations of the hfs splitting. 

Employing the Dirac equation for the electron in the external field of an
infinitely heavy pointlike nucleus, one obtains ($\hbar=c=1$)
\cite{Bre30,Pyyk73,Zap79}
\begin{equation}
A^{\rm D}_{n\kappa} = \alpha (\alpha Z)^3 \frac{m^2}{m_p} 
\frac{g_I \kappa}{j(j+1)}\frac{[2\kappa(n_r + \gamma) - N]}
{N^4\gamma(4\gamma^2 -1 )} , 
\label{eq2}                                       
\end{equation}
where $\alpha = e^2$ is the fine-structure constant $(e>0)$, $g_I$ is the
nuclear $g$ factor, $n_r =n - |\kappa|$ is the radial quantum number, $n$ is
the principal quantum number, $\kappa=(j+1/2)(-1)^{j+l+1/2}$,
$\gamma=\sqrt{\kappa^2 -(\alpha Z)^2}$, $N=\sqrt{(n_r + \gamma)^2 +(\alpha
Z)^2}$, and $m$ and $m_p$ are the electron and proton masses, respectively.
Because of various nuclear and QED effects, the experimental value for the hfs
constant deviates from its Dirac prediction (\ref{eq2}). To describe the
extended nucleus, one usually employs the following parametrization
\begin{equation}
A^{\rm ext}_{n\kappa} = A^{\rm D}_{n\kappa}(1 - \delta_{n\kappa})
(1 - \epsilon_{n\kappa}) ,
\label{eq3}                                       
\end{equation}
where the corrections $\delta_{n\kappa}$ and $\epsilon_{n\kappa}$ account
for the nuclear charge and magnetic moment distributions within the nucleus,
respectively \cite{RoBr32,CrSch49,BoWe50,Shab94}. In addition to the
approximation~(\ref{eq3}), the radiative corrections should be taken into
account. All leading QED effects for the hyperfine structure have been
independently calculated by different groups and the numerical results are
consistent ~\cite{ThBe0}. For Li-like ions, one can develop a perturbation
theory with respect to the parameter $1/Z$, which accounts for corrections
arising from the electron-electron interaction \cite{Sh00,ShSh95}. The recoil
correction, which is due to the finite nuclear mass, is negligibly small for
heavy ions.

Here we consider a correction $\Delta A_{n\kappa}$ to the hfs constant 
(\ref{eq3}) due to the nuclear-polarization effect, which is caused by the
bound electron. More precisely, a core-polarization part of the effect is
considered only, which is due to collective nuclear excitations. The
corresponding single-nucleon contributions should be generally considered
beyond the external-field approximation. However, they are assumed to be
completely negligible. To describe nuclear polarization, we adopt a
relativistic field theoretical approach, which incorporates the many-body
theory for virtual nuclear excitations within bound-state QED for atomic
electrons \cite{GPBM91}. This approach has been successfully applied in
calculations of nuclear-polarization effects to the Lamb shift
\cite{GPBM91,GPS95,NLP96} and to the bound-electron $g$ factor \cite{Nef02}.
To some extent, the present formulae are quite similar to those derived in
Ref.~\cite{Nef02}. The correction under consideration may be represented by
the sum of contributions, which are referred to as the irreducible, the
reducible, and the vertex parts.

The irreducible contribution to the hfs constant can be written in terms of
a multipole decomposition as follows:
\widetext
\begin{equation}
\Delta A_{n\kappa}^{\rm irr} = \frac{\alpha}{2\pi}
\frac{e g_I \mu_N \kappa}{j(j+1)} \sum_{L \ge 0} B(EL) \sum_{n_1,\kappa_1}
\left[ C^{j_1 \frac{1}{2}}_{j \frac{1}{2} L  0} \right]^2  
\frac{\langle n\kappa |F_L|n_1\kappa_1\rangle \langle n_1\kappa_1
|F_L|\overline{n\kappa} \rangle} {\varepsilon_{n\kappa} -
\varepsilon_{n_1\kappa_1} - {\rm sgn}(\varepsilon_{n_1\kappa_1})\omega_L} , 
\label{eq4}                                                 
\end{equation} 
\narrowtext
\noindent
where $\mu_N = e/(2m_p)$ is the nuclear magneton, $\varepsilon_{n\kappa}$
is the one-electron energy, $\omega_L = E_L - E_0$ are the nuclear 
excitation energies with respect to the ground-state energy $E_0$ of the
nucleus and $B(EL)= B(EL;0\to L)$ are the corresponding reduced electric
transition probabilities. The sum over $n_1$ runs over the entire Dirac
spectrum, while the sum over $\kappa_1$ is restricted to those intermediate
states, where $l+l_1+L$ is even. A two-component radial vector 
$\langle r | n\kappa \rangle$ is determined by 
\begin{equation}
\langle r | n\kappa \rangle={ {P_{n\kappa}(r)} \choose {Q_{n\kappa}(r)} }  ,
\label{eq5}                                                    
\end{equation} 
where $P_{n\kappa}(r) =rg_{n\kappa}(r)$ and $Q_{n\kappa}(r) =rf_{n\kappa}(r)$, 
with $g_{n\kappa}(r)$ and $f_{n\kappa}(r)$ being the upper and lower
components of the Dirac wave function, respectively. The radial
shape parametrizing the nuclear transitions is carried by the functions
\cite{GPBM91,GPS95,NLP96}
\widetext
\begin{equation}
F_L(r)=\frac{4\pi}{(2L+1)R_0^L}\left[\frac{r^L}{R_0^{L+1}}\Theta(R_0 -r) + 
\frac{R_0^L}{r^{L+1}}\Theta(r-R_0) \right]  
\label{eq6}                                                  
\end{equation}
\narrowtext
\noindent
in the case of multipole excitations with $L \ge 1$ and
\begin{equation}
F_0(r) =\frac{5\sqrt{\pi}}{2R_0^3}\left[1 -\left( \frac{r}{R_0} \right)^2
\right] \Theta(R_0 -r) 
\label{eq7}                                                  
\end{equation}
for monopole excitations, respectively. Here $R_0$ is an average radius
assigned to the nucleus in its ground state. In Eq.~(\ref{eq4}), the matrix
element is given by
\begin{equation}
\langle a |F_L |b \rangle =\int \limits_0^\infty dr  F_L(r)
\left[P_a(r)P_b(r) + Q_a(r)Q_b(r)\right] .
\label{eq8}                                                   
\end{equation} 
The perturbed vector $\langle r |\overline {n\kappa }\rangle$, which follows
as  
\begin{equation}
\langle r |\overline {n\kappa }\rangle =  \sum_{n'}^{n'\neq n}
\frac{\langle n' \kappa |\sigma_x r^{-2} | n\kappa \rangle} 
{\varepsilon_{n\kappa} - \varepsilon_{n'\kappa}} \langle r|n'\kappa \rangle , 
\label{eq9}                                                   
\end{equation}  
can be evaluated analytically by means of the generalized virial relations for
the Dirac equation \cite{Shab91} (see also Refs.~\cite{ShSh95,ShY96}). In
Eq.~(\ref{eq9}), $\sigma_x$ is the Pauli matrix.

The reducible contribution reads 
\widetext
\begin{equation}
\Delta A^{\rm red}_{n\kappa} = -\frac{\alpha}{4\pi} A^{\rm ext}_{n\kappa}
\sum_{L \ge 0}  B(EL) \sum_{n_1, \kappa_1}  \left[C^{j_1 \frac{1}{2}}_{j
\frac{1}{2} L 0}\right]^2 
\frac{ \langle n\kappa |F_L|n_1\kappa_1\rangle{}^2}
{[\varepsilon_{n\kappa} - \varepsilon_{n_1\kappa_1}  -{\rm 
sgn}(\varepsilon_{n_1\kappa_1})\omega_L]^2} ,
\label{eq10}                                                     
\end{equation} 
\narrowtext
\noindent
where $A^{\rm ext}_{n\kappa}$ is the hfs constant given by Eq.~(\ref{eq3}).
The sum  $l +l_1 + L$ again should be even. 

The nuclear core-polarization correction to the hfs constant due to the vertex
part is conveniently represented as the sum of a pole term
\widetext
\begin{eqnarray}
\Delta A^{\rm pol}_{n\kappa}&=&  \frac{\alpha}{4\pi} \frac{e g_I \mu_N 
\kappa}{\sqrt{j(j+1)(2j+1)}} \sum_{L \ge 0}  B(EL) \sum_{n_1, \kappa_1}
\frac{(2j_1+1)^{3/2}}{\sqrt{j_1(j_1+1)}} \left[C^{j_1 \frac{1}{2}}_{j
\frac{1}{2} L 0}\right]^2 \left\{ {j_1 \atop j} {j_1 \atop j} {1 \atop L}
\right\} \nonumber \\ 
&\times & \frac{\langle n_1\kappa_1 |\sigma_x r^{-2}| n_1\kappa_1 \rangle 
\langle n_1 \kappa_1 |F_L|n \kappa \rangle{}^2}
{[\varepsilon_{n\kappa} - \varepsilon_{n_1\kappa_1}   - {\rm
sgn}(\varepsilon_{n_1\kappa_1}) \omega_L]^2}   
\label{eq11}                                                   
\end{eqnarray}
\narrowtext
\noindent
and of a residual term 
\widetext
\begin{eqnarray}
\Delta A^{\rm res}_{n\kappa}&=& \frac{\alpha}{\pi} \frac{\sqrt{2} e g_I \mu_N 
\kappa}{\sqrt{j(j+1)(2j+1)}} \sum_{L \ge 0}  B(EL) \mathop{{\sum}'}_{n_1,n_2}
\sum_{\kappa_1, \kappa_2} \sqrt{2j_2 + 1}C^{j_1 \frac{1}{2}}_{j \frac{1}{2} L
0}C^{j_2 \frac{1}{2}}_{j \frac{1}{2} L 0} C^{j_1 \frac{1}{2}}_{j_2
-\frac{1}{2}  1 1 } \nonumber \\ 
&\times & \left\{ {j_1 \atop j} {j_2 \atop j} {1 \atop L} \right\} 
\frac{\langle n\kappa |F_L|n_1\kappa_1\rangle \langle n_2\kappa_2 |F_L|
n\kappa\rangle} {\varepsilon_{n\kappa} - \varepsilon_{n_2\kappa_2} - {\rm
sgn}(\varepsilon_{n_2\kappa_2})\omega_L} 
\frac{\langle n_1 \kappa_1 |\sigma_x
r^{-2} | n_2\kappa_2 \rangle}{\varepsilon_{n_1\kappa_1} -
\varepsilon_{n_2\kappa_2}}  ,
\label{eq12}                                                
\end{eqnarray}
\narrowtext
\noindent
respectively. Here $\Delta A^{\rm pol}_{n\kappa}$ accounts for the terms 
with $n_1 = n_2$ and  $\kappa_1=\kappa_2$ in the sums over
intermediate states. The prime in the sum in Eq.~(\ref{eq12}) indicates that
$\varepsilon_{n_1\kappa_1} \neq \varepsilon_{n_2\kappa_2}$ when $\kappa_1
=\kappa_2$, i.e., the pole contribution is supposed to be omitted. In
Eqs.~(\ref{eq11}) and (\ref{eq12}), the value  $l+l_1 + L$ has to be even. A 
second condition in Eq.~(\ref{eq12}) is that the sum $l_1 + l_2$ should be even
as well. The total nuclear core-polarization contribution to the hfs constant
is determined by the sum of all contributions given by Eqs.~(\ref{eq4}), 
(\ref{eq10}), (\ref{eq11}), and (\ref{eq12}).

In Table~\ref{table1}, we present numerical results for some hydrogenlike ions,
which are of particular experimental interest. The calculations were
performed taking into account a finite set of dominant collective nuclear
excitations. To estimate the nuclear parameters, $\omega_L$ and $B(EL)$,
in the case of nearly spherical nucleus of $^{209}_{\ 83}$Bi, we employed
experimental data corresponding to the low-lying vibrational levels in
neighbouring even-even isotope of $^{208}_{\ 82}$Pb, which were deduced from
nuclear Coulomb excitation. In the case of giant resonances, we utilized
phenomenological energy-weighted sum rules \cite{RS78}, which are assumed to
be concentrated in single resonant states. In the present calculations,
contributions due to monopole, dipole, quadrupole, and octupole giant
resonances have been taken into account. The infinite summations over the
entire Dirac spectrum were performed by the finite basis set method. Basis
functions are generated via B splines including nuclear-size effects
\cite{JBS88}.

Concluding, we have evaluated a correction to the hyperfine structure in heavy
multicharged ions, which is connected with the nuclear core-polarization
effect caused by the unpaired bound electron. The correction is exhausted
over distances of the order of the nuclear size and it is enhanced due to a
singular behavior of the hyperfine-interaction operator. The uncertainty of
our calculation can be as large as the nuclear-polarization effect itself.  
It yet determines the natural limitation for testing higher-order QED
corrections in future experiments aiming for accurate hfs measurements. In the
case that the experimental value for the ground-state hfs splitting is used to
eliminate the Bohr-Weisskopf effect \cite{Shab98}, the latter cannot be
separated from the effect we have considered. Accordingly, the utmost
precision for theoretical hfs predictions in lithiumlike heavy ions is
determined by the nuclear-polarization correction to the ground state in 
hydrogenlike ions. In particular, for the ground state in $^{209}$Bi$^{82+}$,
the nuclear-polarization effect contributes on the level of about $0.05$ meV.
This implies that although numerical calculations of bound-state QED
corrections provide sufficiently stable results, the conservative estimate of
uncertainties for the hfs splitting in $^{209}$Bi$^{80+}$ quoted in
Ref.~\cite{Sh00} appears to be more realistic rather then the one predicted in
Ref.~\cite{SSc01}. It is also worth noting that because of a similar scaling
dependence of nuclear and radiative corrections upon the principal quantum
number, significant cancellations of almost all corrections, except for the
screened QED contribution, occur in a specific difference of the hfs
splittings in H- and Li-like heavy ions \cite{ShAN01}. In this case, the
nuclear-polarization effect might be most relevant for determining the 
uncertainties of accurate theoretical predictions.

\acknowledgments

A.N. is grateful for financial support from RFBR (Grant No. 01-02-17246) and
from the Alexander von Humboldt Foundation. G.S. and G.P acknowledge financial
support from BMBF, DFG, and GSI.

\newpage
\begin{table}[h]
\widetext
\noindent
\caption{For various hydrogenlike ions, the nuclear spin/parities $I^\pi$,
the nuclear magnetic dipole moments $\mu$ (in units of the nuclear magneton)
[29], the nuclear-polarization contributions to the hfs constant $\Delta
A_{1s}$ for K-shell electron, and the nuclear-polarization corrections to the
ground-state hfs splitting $\Delta E_{1s}$ are tabulated. Column (a):
contributions from low-lying collective nuclear modes using experimental
values for nuclear excitation energies $\omega_L$ and electric transition
strengths $B(EL)$; (b) contributions from giant resonances employing empirical
sum rules [30]; (c) total effect. The negative value of the nuclear magnetic
moment for uranium indicates that the level with $F=I-1/2$ lies above the one 
with $F=I+1/2$. The numbers in parentheses are powers of ten.}
\label{table1}                         
\end{table}
\noindent
\begin{tabular}{llr@{}lr@{}lr@{}lr@{}lc} \hline 
\multicolumn{1}{c}{ } & \multicolumn{1}{c}{$I^\pi$} &
\multicolumn{2}{c}{$\mu$ (nm)} & \multicolumn{6}{c}{$\Delta A_{1s}$ (meV)} &
\multicolumn{1}{c}{$\Delta E_{1s}$ (meV)} \\ 
\hline
$^{159}$Tb$^{64+}$ & ${3/2}^+$ & 2 & $.014$ & 0 & $.25(-2)^{\rm  a}$ &
0 & $.26(-2)^{\rm  b}$ & 0 & $.51(-2)^{\rm  c}$ & $0.10(-1)$ \\ 
$^{165}$Ho$^{66+}$ & ${7/2}^-$ & 4 & $.132$ & 0 & $.38(-3)$ &
0 & $.27(-2)$ & 0 & $.31(-2)$ & $0.12(-1)$ \\ 
$^{175}$Lu$^{70+}$ & ${7/2}^+$ & 2 & $.2327$ & 0 & $.13(-2)$ &
0 & $.22(-2)$ & 0 & $.35(-2)$ & $0.14(-1)$ \\ 
$^{187}$Re$^{74+}$ & ${5/2}^+$ & 3 & $.2197$ & 0 & $.25(-2)$ &
0 & $.66(-2)$ & 0 & $.91(-2)$ & $0.27(-1)$ \\ 
$^{203}$Tl$^{80+}$ & ${1/2}^+$ & 1 & $.62226$ & 0 & $.16(-2)$ &
0 & $.29(-1)$ & 0 & $.31(-1)$ & $0.31(-1)$ \\ 
$^{209}$Bi$^{82+}$ & ${9/2}^-$ & 4 & $.1106$ & 0 & $.81(-3)$ &
0 & $.10(-1)$ & 0 & $.11(-1)$ & $0.55(-1)$ \\ 
$^{235}$U$^{91+}$ & ${7/2}^-$ & $-0$ & $.38$ & $-0$ & $.36(-2)$ &
$-0$ & $.29(-2)$ & $-0$ & $.65(-2)$ & $0.26(-1)$ \\ 
\hline 

\end{tabular}

\end{document}